# How Does Ontology Contribute in Semantic Web Development?

Zeeshan Ahmed and Detlef Gerhard

Mechanical Engineering Informatics and Virtual Product Development Division (MIVP),
Vienna University of Technology,
Getreidemarkt 9/307 1060 Vienna, Austria
{zeeshan.ahmed, detlef.gerhard}@tuwien.ac.at

**ABSTRACT**
This paper investigates and briefly describes the major currently existing problems with World Wide Web .i.e., Information filtration and Security became the main reasons of semantic web's invention. The semantic web claims of providing the semantic based solutions towards current web problems.  Semantic web have introduced and relies on a main building block "Ontology" to provide the information in machine processable semantic models and produce semantically modelled knowledge representation systems.  This paper also describes the role, construction process and the contributions of ontology in providing some in time proposed and implemented solutions. Furthermore paper concludes with the currently existing limitations in Ontology and the areas which need improvements.

## 1. INTRODUCTION

World Wide Web is an automatic and cheapest global information sharing and communication system made up of three standards Uniform Resource Identifier (URL), Hypertext Transfer Protocol (HTTP) and Hypertext Mark-up Language (HTML) by Tim Berners-Lee to effectively store, communicate and share different forms of information. The information is provided over the web in text, image, audio and video formats using HTML, considered unconventional in defining and formalizing the meaning of the context.  Most of the information is structured only inside the available databases over the web and due to this it is quite easy to go for scattered extensive information by looking into bookmarked web pages but quite difficult to extract a piece of needed information. Although some search engines and screen scrapers are invented, search engine uses full text query to search information but can only return unstructured contents not the actual structured information stored in database on web where as screen scrapers extracts and repurpose fragments from web pages but insufficient in creating a rich multi domain information environment [3]. Most of the search engines are not satisfactory because they requires excessive manual preprocessing e.g. designing a schema, cleaning raw data, manually classifying documents into a taxonomy and manual post processing e.g. browsing through large result lists with too many irrelevant items [7]. To increase the integration and interoperability over the web the concept of "Web Service" was introduced. Due to the dynamic nature web services became very famous in industry in short time but with the passage of time due to the heavily increase in number of web services end-to-end service authentication, authorization, data integrity and confidentiality problems were identified which are still alive and not handled by existing web technologies [8].

To cope with the currently existing web problems .i.e., Information filtration, security, confidentiality and augmentation of meaningful contents in mark-up presentation over the web a semantic based solution "Semantic Web" was proposed by Tim Berners Lee [19]. The semantic web is an intelligent incarnation and advancement in World Wide Web to collect, manipulate and annotate information independently by providing effective access to the information. Semantic web provides categorization and uniform access to resources, promotes the transformation of World Wide Web in to semantically modelled knowledge representation systems and common framework which allows data to be shared and reused [21]. Semantic web also gives the concept of semantic based web services to provide solutions to the problems of dynamically composed service based applications. Moreover semantic web aims of providing information in machine processable semantic models which assigns information resources to classes whose meaning is defined in Ontologies [9], a collection of interrelated semantic based concepts briefly described in section 2.

## 2. ONTOLOGY

Ontology is playing a vital role in solving the existing web problems by producing semantic aware solutions. Ontology makes machines capable of understanding the semantic languages that humans use and understand by producing the abstract modelled representation of already defined finite sets of terms and concepts involved in intelligent information integration and knowledge management [4]. Ontology is basically categorized in three different kinds .i.e., Natural Language Ontology (NLO), Domain Ontology (DO) and Ontology Instance (OI) to provide relationships between generated lexical tokens of statements based on natural language, knowledge of a particular domain and to generate automatic object based web pages [2]. Ontologies are constructed and connected to each other in a decentralized manner to clearly express semantic contents and arrange semantic boundaries to find out required needed information [16].

### 2.1 Ontology Construction

Ontology construction is a highly relevant research issue depends on the extraction of information from web and emergence of ontologies. Natural language based information is treated as the input to the ontology construction process, which parses the text in nouns and verbs. Nouns are represented as "Classes" and verbs as "Properties" containing values, relationships with other properties and some constraints. Classes are further divided in main and sub class categories maintained in taxonomy hierarchy. The size of ontology varies due to the increase in number of classes and instances.

First step in building ontologies is to create the nodes and edges, once the concepts and relationships of a graph based ontology are constructed then next step is to quantify the strengths of semantic relationships [7]. Ontologies can be constructed manually and automatically by using some ontology supporting languages .i.e., XML (eXtensible Mark-up Language) [18], RDF (Resource Description Framework) [17] and OWL (Web Ontology Language) [5] offering ways of more explicitly structuring and richly annotating Web pages.

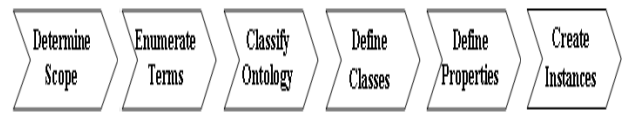

Fig 1. Iterative ontology development process

Ontology development is an iterative process based on six main activities .i.e., Determine Scope, Enumerate Terms, Classify Ontology, Define Classes, Define Properties and Create Instances as shown in Fig 1. In the beginning of ontology development process it is very important to determine the scope otherwise it will be very time and effort consuming. Then enumerated terms should be identified to classify ontologies with in their respective types. Classes and their respective properties along with their relationships and constraints are defined using identified enumerated terms. In the end only the instances are created and used. To implement ontology development process professional experience, powerful user friendly ontology supporting tools and communication between domain experts and developers is required [10].

### 2.2 In Time Ontology based Solutions

Ontology heavily contributed in the industry by supporting the development of advanced language processing tools and very large linguistics resources [6]. Now a days Ontology is becoming very popular especially in semantic web and desktop communities; professionals are using RDF and OWL extensively to take advantage in providing solutions to the problems of Information filtration, extraction and security over the web. In time some ontology based solutions are already proposed and developed including Semantic Desktop [14], Reisewissen [12], Test Application [15], Semantic Security Web Services (SSWS) [8], Cultural Heritage and the Semantic Web [20].

### 2.2.1 Semantic Desktop

Personal Information Model [14]. Semantic Desktop is proposed to improve the process for the identification of documents and retrieval of no unnecessary document. The designed architecture mainly consists of three main components .i.e., Receiver, Interpreter and Analyzer which provides index services, structuring of the contents with the help of manual annotation and Meta data. The development of Semantic Desktop is totally based on ontology and classes, the relationships of classes and ontologies are predefined and the information is accessed using RDF graphs.

Semantic Security Web Services (SSWS) [8]. SSWS is an ontology based solution towards end-to-end service authentication, authorization, and data integrity and confidentiality problems introduced by web services. SSWS is capable of improving currently available concept of web services by adding semantically meaningful, declarative and machine process able descriptions of security. SSWS is developed using semantic web technologies OWL, XML, context of DARPA Agent Mark-up Language (DAML) and some existing web's supporting technologies SOAP, WSTK, Apache Tomcat Web, HTML, JSP, and Servlets.

### 2.2.2 Reisewissen [12]

Reisewissen is proposed to identify potential relevant knowledge sources and provide quality services by semantically connecting, organizing and sharing the currently isolated pieces of information in an online portal to anticipating customer behavior. The design of Reisewissen is composed of three main components .i.e., Data Connectors (DC), Evaluation Framework (EF) and Evaluation Engine (EE) which provides data sources, transformation of data from heterogeneous to common data format, caching and fetching of data, a workbench to test the quality of data and ranking of information. Reisewissen is implemented using Ontology in a real time application, a hotel recommendation engine and travel information system. Information is obtained using Simple Object Access Protocol (SOAP) based web services and stored in both RDF and non RDF formats, which then matched to find out the desired result. Data stored in RDF format is based on developed ontologies mapped between database and RDF triples. Data is matched semantically by combining data properties to Ontology and similarities between two concepts are determined by distance reflecting their respective positions in hierarchy and as the result list of selected results are generated to customer.

### 2.2.3 Cultural Heritage and the Semantic Web [20]

Cultural Heritage and the Semantic Web is ontology based proposed approach and implemented in a tool to semantically annotate existing cultural contents by supporting the annotation process by an intelligent editor. Proposed approach is limited to some extent but still provides the process of semantic navigation, intelligent search, 3D visualization and methodology to publish and exploit content on the Semantic Web.

### 2.2.4 Test Application [15]

Test Application is proposed to publish information on web without inserting into relational database by making a flexible reasoning system and to take advantage in improving already existing semantic search mechanism is proposed. The approach implemented as Test Application using Ontology, each web object is referred as Unified Resource Identifier (URI) and to provide context in machine interpretable way every key principle is implemented in RDF and OWL format. Information is first converted in to ontologies, then using a translator application transformed into machine understandable format and then saved in to RDF files. These RDF files can easily be published on web as well as retrieved by any search engine based on digital assistants.

## 2.3 Limitations in Ontology

Now days the development of ontology driven applications is slowed down due to some limitations and principal problems which are [22]

Existing natural language parsers used to parse the information to construct ontologies are limited because they can only work over a single statement at a time [6].

Existing methodologies of structuring ontologies are in sufficient and need to be improved because now it is quite impossible to define the boundaries of ontology based on particular domain's abstract model and automatically handle the increase in size of ontology due to the increase in number of classes and instances.

Creating ontologies manually is a time consuming process which becomes very complex when there is a large amount of data to create large number of ontologies. To take advantage in creating large number of ontologies by reducing the complexity and time an automatic ontology creation mechanism is required. Some mechanisms are already proposed and implemented to create ontologies automatically but those are insufficient and less qualitative. While creating nouns based classes using existing automatic ontology creation mechanism, it is quite impossible to identify the possible existing relationships between classes to draw the taxonomy hierarchy [11]. Furthermore it is also quite impossible to perform automatic emergence of ontologies to create new ontologies [13].

Currently available ontology validators are restricted and not capable of validating all kind of ontologies e.g. based on complex inheritance relationship [8].

Domain specific ontologies are highly dependent on the domain of the application and because of this dependency domain specific ontology's contained specific senses are not possible to find in general purpose ontology [1].

The process of semantic enrichment reengineering for the web development consists of relational metadata required to be developed at high speed and in low cost depending on proliferation of ontologies, which is currently also not possible.

Handling of dynamically raised calculations caused by the comparison of big complexities of similar ontologies is also not possible (Marc et al. 2004).

## 3. CONCLUSION

In this paper we have discussed information filtration and security problems with web become the major causes to improve the concept of web in to semantic web. We have briefly described semantic web and how using ontology it is contributing in providing the solutions towards current web problems.

We have discussed ontology as the major role player in providing semantic based solutions to the software industry by contributing in real time software applications e.g. helps in identifying documents and retrieval of no unnecessary document, provides solution towards end-to-end service authentication, authorization, data integrity and confidentiality, helps in identifying potential relevant knowledge sources, provides quality services by semantically connecting, organizing and sharing the currently isolated pieces of information, helps in publishing information on web without inserting into relational database by making a flexible reasoning system. Furthermore we have also shortly discussed how ontology can be constructed using in time available development processes and supporting development languages. No doubt semantic web using ontology have contributed the in progress of web but still there are some limitations and due to those semantic web is currently not succeeded in attaining the actual goal of completely structuring the information over the web which can be processed by machines and making advanced knowledge modelled system. In future, the need is to enhance the concept of ontology with respect to implementation point of view because all the theories can be fruitful if the implementation is possible.

## 4. REFERENCES

[1] Amalia, T, Laurent, R & Dalila. B., 2002. "Vulcain – An Ontology-Based Information Extraction System", pp. 64–75. Springer-Verlag B. Andersson et al. (Eds.): NLDB, LNCS 2553, Berlin Heidelberg Germany

[2] Borys, O 2001, "Learning of ontologies for the Web: the analysis of existent approaches", *In Proceedings of the International Workshop on Web Dynamics*, held in conj. with the 8th International Conference on Database Theory (ICDT'01), London, UK

[3] David, H, Stefano, M, & David, K, 2007, "Piggy Bank: Experience the Semantic Web Inside Your Web Browser", pp. 413 – 430, Y. Gil et al. (Eds.): ISWC, LNCS 3729, Amsterdam Netherlands

[4] Dieter. F., 2000, "Ontologies: Silver Bullet for Knowledge Management and Electronic Commerce". Springer-Verlag, Berlin Germany

[5] Deborah, L, McGuinness & Frank, V Harmelen, 2004 OWL Web Ontology Language, Retrieved 15 May 2007 from http://www.w3.org/TR/owl-features/

[6] Galia, A 2005, "Language Technologies Meet Ontology Acquisition", pp. 367-380, Springer-Verlag F. Dau, M.-L. Mugnier, G. Stumme

(Eds.): ICCS, LNAI 3596, Berlin Heidelberg Germany

[7] Gerhard, W, Jens .G, Ralf .S & Martin .T, 2004, "Towards a Statistically Semantic Web", pp. 3–17, Springer-Verlag P. Atzeni et al. (Eds.): ER, LNCS 3288, Berlin Heidelberg Germany

[8] Grit, Dr, Son, N, & Andrew, T, 2004, "OWL-S Semantics of Security Web Services: a Case Study", pp. 240–253, Springer-Verlag J. Davies et al. (Eds.): ESWS, LNCS 3053, Berlin Heidelberg Germany

[9] Heiner, S 2002, "Approximate Information Filtering on the Semantic Web", pp. 114–128, Springer-Verlag M. Jarke et al. (Eds.): KI, LNAI 2479, , Berlin Heidelberg Germany

[10] Holger, K, Mark, A. Musen & Natasha F. Noy, 2003, "Tutorial: Creating Semantic Web (OWL) Ontologies with Protégé", *In proceedings of 2nd International Semantic Web Conference*, October, Sanibel Island, Florida, USA

[11] Jos´e, S & Paulo, Q, 2005, "A Methodology to Create Legal Ontologies in a Logic Programming Information Retrieval System", pp. 185–200, Springer-Verlag V.R. Benjamins et al. (Eds.): Law and the Semantic Web, 3369, Berlin Heidelberg Germany

[12] Magnus, N, Malgorzata, M & Robert, T, 2006, "Improving Online Hotel Search What Do We Need Semantic For", In Proceedings of Semantic Systems from Visions to Applications, Vienna Austria

[13] Marc, E & York Sure, 2004, "Ontology Mapping - An Integrated Approach", pp. 76–91, Springer-Verlag J. Davies et al. (Eds.): ESWS, LNCS 3053, Berlin Heidelberg Germany

[14] Mark, S, Pierre, S, Leo, S & Andreas, D, 2006, "Increasing Search Quality with the Semantic Desktop in Proposal Development", *In the proceedings of Practical Aspects of Knowledge Management 6th International Conference.* (PAKM), Vienna Austria

[15] Markus, L, Martin, S, Dieter, F, Schahram, D, Heinrich, O & Tassilo, P., 2006, "The realization of Semantic Web based E-Commerce and its impact on Business, Consumers and the Economy", *In Proceedings of Semantic Systems From Visions to Applications*, Vienna Austria

[16] Okkyung, C, SeokHyun, Y, Myeongeun, O, & Sangyong, H, 2003, "Semantic Web Search Model for Information Retrieval of the Semantic Data", pp. 588-593,Springer-Verlag C.-W. Chung et al. (Eds.): HSI LNCS 2713, Berlin Heidelberg Germany

[17] Sean B. Palmer 2007, The Semantic Web: An Introduction, Retrieved May 10, 2007 from http://infomesh.net/2001/swintro/

[18] Liam, Q 2007, The W3C Extensible Markup Language (XML), Retrieved May 10, 2007 from http://www.w3.org/XML

[19] Tim, B. Lee, James, H & Ora, L, 2001, The Semantic Web, A new form of Web content that is meaningful to computers will unleash a revolution of new possibilities, Retrieved June 30, 2007 from http://www.geodise.org/useful_links/link_semantic.htm

[20] V.R, Benjamins, J. Contreras, M. Blázquez, J.M. Dodero, A. Garcia, E. Navas, F. Hernandez & C. Wert, 2004, "Cultural Heritage and the Semantic Web", pp. 433-444, Springer-Verlag J. Davies et al. (Eds.): ESWS LNCS 3053, Heidelberg Germany

[21] Witold, A, Tomasz, K & Krzysztof, W, 2005, "How Much Intelligence in the SemanticWeb?", pp. 1−6, Springer Verlag P.S. Szczepaniak et al. (Eds.): AWIC, LNAI 3528, Berlin Heidelberg Germay

[22] Zeeshan. A, Detlef. G, "Role of Ontology in Semantic Web Development", In First International Workshop on Cultural Heritage on the Semantic Web in conjunction with the 6th International Semantic Web Conference and the 2nd Asian Semantic Web Conference, November 2007, Busan Korea